\documentclass[twocolumn,floatfix,prd,authoryear,nofootinbib,aps,10pt,tightlines,eqsecnum]{revtex4-1}
\usepackage{bm}
\usepackage{newtxmath, newtxtext}
\usepackage{graphicx}
\usepackage{textcomp}
\usepackage{float}
\usepackage{booktabs}
\usepackage{dcolumn,enumerate}
\usepackage{ragged2e}
\usepackage{hyperref}

\hypersetup{colorlinks=true, linkcolor=blue, urlcolor=blue, citecolor=blue}

\usepackage{xcolor}
\usepackage{epsfig}
\usepackage{caption}
\usepackage{appendix}
\usepackage{subcaption}
\usepackage{graphicx} 
\usepackage{orcidlink}
\begin{document}

\title{Exploring the accelerating black holes from the observations of quasi-periodic oscillations in X-ray binaries}

\author{Hamza Rehman $^{a, b, c}$}
\email{hamzarehman244@zjut.edu.cn}
\author{Saddam Hussain$^{a, b}$ \orcidlink{0000-0001-6173-6140}}
\email{saddamh@zjut.edu.cn}
\author{G. Abbas ${}^{d}$}
\email{ ghulamabbas@iub.edu.pk}
\author{Tao Zhu $^{a, b}$ \orcidlink{0000-0003-2286-9009}}
\email{Corresponding author: zhut05@zjut.edu.cn}

\affiliation{${}^{a}$ Institute for Theoretical Physics and Cosmology, Zhejiang University of Technology, Hangzhou 310023, China}
\affiliation{${}^{b}$ United Center for Gravitational Wave Physics (UCGWP), Zhejiang University of Technology, Hangzhou, 310023, China}
\affiliation{${}^{c}$ Center for Theoretical Physics, Khazar University, 41 Mehseti Str., Baku, AZ1096, Azerbaijan}
\affiliation{${}^{d}$ Department of Mathematics, The Islamia University of Bahawalpur, Bahawalpur, Pakistan}

\date{\today}

\begin{abstract}

Black holes in dense astrophysical environments, such as globular clusters or in the vicinity of other massive objects, may possess accelerations. Such acceleration would modulate the characteristics of the quasi-periodic oscillations (QPOs) observed in X-ray black hole binaries. In this paper, we explore the influence of spin-aligned acceleration of a black hole on QPOs observed in X-ray binaries. For this purpose, we compute the fundamental frequencies arising from the motion of test particles around an accelerating (spin-aligned) black hole and apply the relativistic precession, parametric resonance, and forced resonance models to establish their correspondence with several observed QPOs of X-ray binaries (GRO J1655-40, XTE J1550-564, XTE J1859+226, GRS 1915+105, H1743-322, M82~X-1, and Sgr~A$^{*}$). We then employ the Bayesian Markov-Chain Monte Carlo method to constrain the black hole parameters. Our results show no evidence for spin-aligned acceleration in any of the analyzed sources, suggesting that most of these X-ray binaries reside in isolated environments and therefore experience only small perturbations to the background spacetime geometries.\\

\end{abstract}

\maketitle

\section{Introduction}

General Relativity (GR) speculates about the existence of black holes (BHs), offering insights into gravity and spacetime. The study of BHs improves our understanding of gravity and delineates the boundaries of physics by manifesting behaviors that bridge quantum and classical theories. Furthermore, a benchmark validation of GR was achieved through the detection of gravitational waves originating from binary BH mergers by the LIGO-Virgo collaboration \cite{LIGOScientific:2016aoc}. Subsequently, the Event Horizon Telescope (EHT) captured unprecedented images of supermassive BHs, specifically M87* and SgrA*, which are located at the center of the M87 and Milky Way galaxies \cite{EventHorizonTelescope:2019dse, EventHorizonTelescope:2019uob, EventHorizonTelescope:2019jan, EventHorizonTelescope:2019ths, EventHorizonTelescope:2019ggy, EventHorizonTelescope:2019pgp}. These groundbreaking accomplishments sparked a new era of astronomy and solidified BHs as significant astrophysical phenomena. Apart from direct imaging, X-ray binaries' quasi-periodic oscillations (QPOs) offer an effective way for examining the geometry of the space surrounding the BH and the nature of gravity in the strong field.

QPOs are periodic variations in the X-ray intensity of an accreting compact star system, particularly observed in X-ray binaries. They were first investigated in the 1980s \cite{Samimi:1979si} as an intriguing astrophysical phenomenon associated with the relativistic motion of matter accreting onto compact objects through an accretion disk. High-resolution timing of X-ray oscillations in X-ray binaries provides a powerful probe of the spacetime geometry and strong-field gravity in the immediate vicinity of compact objects \cite{Stella:1998mq, Stella:1997tc}. In an X-ray binary, the compact object (a BH or a neutron star) accretes matter from a stellar companion; the inflowing gas forms an accretion disk whose inner regions emit X-rays. Temporal features in these X-rays, in particular QPOs, encode properties of both the gravitational field and the accretion flow and hence can resolve spatial scales far below the limits of current imaging techniques \cite{Ingram:2019mna, Remillard:2006fc}.

Numerous theoretical models have been proposed to explain the QPO phenomenon, such as the relativistic precession model (RP), epicyclic resonance (ER), forced resonance models (FR), warped disk (WD) model, the parametric resonance models (PR) \cite{Stella:1997tc, Stella:1998mq, 1999ApJ...524L..63S, Cadez:2008iv, Kostic:2009hp, Germana:2009ce, Kluzniak:2002bb, Abramowicz:2003xy, Rebusco:2004ba, Nowak:1996hg, Torok:2010rk, Torok:2011qy, Kotrlova:2020pqy}. These models relate the observed timing features to orbital motion and small perturbations of test-particle trajectories in strong gravity, and therefore make QPOs an effective tool for testing relativistic dynamics near compact objects. In this work, we concentrate on three representative prescriptions widely used in the literature: the RP model (linking QPOs to orbital, radial, and vertical epicyclic frequencies), the PR model (describing nonlinear coupling and parametric resonances between radial and vertical oscillations), and the FR model (where resonant response is driven by disk or external perturbations). Together, these models capture the principal mechanisms by which the accretion flow can produce the characteristic QPO frequency ratios and amplitudes observed in X-ray binaries.

Observed QPOs originate from gas orbiting close to the compact object and therefore carry direct information about strong-field relativistic effects. Although many early studies focused on high-frequency QPOs in neutron-star systems, related models have been extended to both stellar-mass and supermassive BHs \cite{PhysRevLett.82.17}. BHs provide a relatively ``clean" astrophysical laboratory for probing spacetime geometry and testing gravity in the strong-field regime \cite{Motta:2013wga}. Accordingly, QPO studies have been applied to tests of the no-hair theorem and to searches for deviations from Kerr geometry in a variety of contexts (e.g., GRO J1655-40 and other BH candidates, non-linear electrodynamics, wormholes, and modified-gravity scenarios) \cite{Allahyari:2021bsq, Banerjee:2022chn, Bambi:2012pa, Bambi:2013fea, Deligianni:2021ecz, Deligianni:2021hwt, Maselli:2014fca, Wang:2021gtd, Jiang:2021ajk, Ashraf:2025lxs, Yang:2025aro, Guo:2025zca, Yang:2024mro, Liu:2023ggz, DeFalco:2023kqy, Bambi:2022dtw, Liu:2023vfh}. The motion of test particles and the resulting epicyclic frequencies in various BH spacetimes have been examined extensively \cite{Dasgupta:2025fuh, Banerjee:2021aln, Jumaniyozov:2025wcs, Borah:2025crf, Rehman:2025hfd, Shaymatov:2023rgb, Stuchlik:2015sno, Banerjee:2022ffu}.

Recent observational and theoretical advances have also motivated consideration of more complex astrophysical formation channels and environmental effects. Black holes that formed or reside in dense environments — for example, in globular clusters, or in the vicinity of other massive bodies — can experience a nonzero net acceleration. Although modeling a generic accelerated, rotating black hole is challenging, acceleration can be incorporated under certain symmetry assumptions. The Kerr-C metric provides one such example \cite{Plebanski:1976gy}: it is an exact vacuum solution of Einstein's field equation describing an accelerating, rotating BH solution with spin-aligned acceleration. Such accelerations can modify observable signatures. : they alter lensing time delays, shift the optimal viewing inclination for shadows, and generally perturb geodesic motion near the BH, see refs.~\cite{Mellor:1989gi, Mann:1995vb, Dias:2003st, Hawking:1995zn, Eardley:1995au, Garfinkle:1990eq, Dowker:1994up, Kinnersley:1970zw, Gussmann:2021mjj, Morris:2017aa9985, Ashoorioon:2022zgu, JahaniPoshteh:2022yei, Zhang:2020xub, Grenzebach:2015oea, EslamPanah:2024dfq, Sui:2023rfh, Zhang:2020xub} and references therein. In particular, accelerated BHs may imprint measurable changes on the timing properties of accreting systems: acceleration can modulate the epicyclic frequencies and thus the characteristics of the QPOs observed in X-ray BH binaries, which was previously explored in ref.~\cite{Sui:2025yem}. Motivated by these considerations, we investigate QPO models in the spacetime of accelerating black holes and quantify how the acceleration parameter affects geodesic motion, epicyclic frequencies, and the resulting observable QPO properties. We consider seven different X-ray binary sources, spanning a range of masses including stellar-mass, intermediate-mass, and supermassive BH systems. To explore the parameter space of these systems, we employ the Bayesian Markov-Chain Monte Carlo method to constrain the black hole parameters.

This article is structured as follows. In Sec.~II, we present a fundamental derivation of the QPO frequencies using the Euler-Lagrange equation of motion for massive particles in an accelerating spacetime. Sec.~III discusses the frequency prescriptions for QPO oscillations, including the RP, PR, and FR models. Sec.~IV analyzes the X-ray QPO observational data and the Markov Chain Monte Carlo (MCMC) analysis. Sec.~V describes the best-fit values obtained from the MCMC simulations used to constrain the BH parameters. Finally, Sec.~VI summarizes our main findings and conclusions.

{\em Note added: While preparing this manuscript, ref.~\cite{Sui:2025yem} appeared, which also investigates QPO signatures of accelerating BH. Our work differs in methodology and interpretation: we perform a Bayesian parameter inference using MCMC techniques, and we adopt physical models for the QPOs that are different from ref.~\cite{Sui:2025yem}. We also consider different X-ray binary sources.}


\section{Mathematical framework of accelerating BH and the corresponding QPOs frequencies}

In Boyer-Lindquist coordinates, the line element of the accelerating BH can be presented as: \cite{Zhang:2020xub} 
\begin{eqnarray}
ds^{2}& = &\frac{1}{\Omega^{2}} \Bigg[ 
\Sigma \left( \frac{d\theta^{2}}{\Delta_{0}} + \frac{dr^{2}}{\Delta_{r}} \right) 
- \frac{\Delta_{r} - a^{2}\Delta_{0}\sin^{2}\theta}{\Sigma} \, dt^{2} \nonumber \\&&
+ \frac{2[ \chi \Delta_{r} - a \Delta_{0}\sin^{2}\theta (a\chi + \Sigma)]}{\Sigma} \, dt d\phi \nonumber \\&&
+ \frac{\Delta_{0}\sin^{2}\theta (a\chi + \Sigma)^{2} - \chi^{2}\Delta_{r}}{\Sigma} \, d\phi^{2} 
\bigg], \label{za1}
\end{eqnarray}
where
\begin{eqnarray}
&&\chi = a \sin^{2}\theta, \  
\Omega = 1 - A r \cos\theta, \
\Sigma = r^{2} + a^{2}\cos^{2}\theta, \nonumber \\&&
\Delta_{r} = (1 - A^{2} r^{2})(r^{2} - 2 m r + a^{2}), \nonumber \\&&
\Delta_{0} = 1 - 2 A m \cos\theta + a^{2} A^{2} \cos^{2}\theta,
\end{eqnarray}
where $m$ is the mass, $a = J/m$ is the angular momentum per unit mass with total angular momentum $J$, and $A$ denotes the BH acceleration. The conformal factor satisfies $\Omega > 0$ and vanishes at the conformal boundary $r_{A} = 1/(A \cos\theta)$. 

In this section, we determine the QPOs around the accelerating Kerr spacetime. To study the QPOs in the accelerating Kerr spacetime, we analyze the geodesic motion of a test particle and derive the fundamental frequencies that characterize its motion in this geometry. The analysis begins with the Lagrangian of the particle.
\begin{equation}
\mathcal{L} = \frac{1}{2} g_{\mu\nu} \frac{dx^\mu}{d\lambda} \frac{dx^\nu}{d\lambda}.\label{za2}
\end{equation}
Here, $\lambda$ is the affine parameter of the particle's worldline. For massless particles, $\mathcal{L}=0$, but for massive particles, $\mathcal{L}<0$. The corresponding generalized momentum is given as
\begin{equation}
p_\mu = \frac{\partial \mathcal{L}}{\partial \dot{x}^\mu} = g_{\mu\nu} \dot{x}^\nu. \label{a1}
\end{equation}
From Eq.~(\ref{a1}), we acquired the equations of motion
\begin{eqnarray}
p_t &=& g_{tt} \dot{t} + g_{t\phi} \dot{\phi} = -\tilde{E}, \label{a2}\\
p_\phi &=& g_{t\phi} \dot{t} + g_{\phi\phi} \dot{\phi} = \tilde{L}, \label{a3}\\
p_r &=& g_{rr} \dot{r}, \\
p_\theta &=& g_{\theta\theta} \dot{\theta}, \label{a4}
\end{eqnarray}
where $\tilde{E}$ is the conserved energy, $\tilde{L}$ represents the conserved angular momentum of the particles, and the overdot denotes the derivative with respect to the affine parameter $\lambda$. From the above equations, we obtained
\begin{eqnarray}
\dot{t} &=& \frac{g_{\phi\phi} \tilde{E} + g_{t\phi} \tilde{L}}{g_{t\phi}^2 - g_{tt} g_{\phi\phi}}, \label{a5}\\
\dot{\phi} &=& \frac{\tilde{E} g_{t\phi} + g_{tt} \tilde{L}}{g_{tt} g_{\phi\phi} - g_{t\phi}^2}. \label{a6}
\end{eqnarray}
By using normalization condition, $g_{\mu \nu} \, \dot{x}^{\mu} \, \dot{x}^{\nu} = -1$ and Eqs.~(\ref{a5}) and (\ref{a6}), we have
\begin{equation}
g_{rr} \, \dot{r}^{2} + g_{\theta\theta} \, \dot{\theta}^{2} = -1 - g_{tt} \, \dot{t}^{2} - g_{\phi\phi} \, \dot{\phi}^{2} - 2 g_{t\phi} \, \dot{t} \, \dot{\phi}. \label{a7}
\end{equation}
For the sake of simplicity, we consider equatorial motion of the particles, i.e., $\theta = \pi/2$, $\dot{\theta} = 0$, and solving Eqs.~(\ref{a5})--(\ref{a7}), we obtain 
\begin{equation}
\dot{r}^{2} = V_{\text{eff}}(r, M, \tilde{E}, \tilde{L}) =
\frac{\tilde{E}^{2} g_{\phi\phi} + 2 \tilde{E} \tilde{L} g_{t\phi} + \tilde{L}^{2} g_{tt}}{g_{t\phi}^{2} - g_{tt} g_{\phi\phi}} - 1. \label{a8}
\end{equation}
Here, $V_{\text{eff}}(r, M, \tilde{E}, \tilde{L})$ is the effective potential for a particle with specific energy $\tilde{E}$ and angular momentum $\tilde{L}$. In the equatorial plane, a stable circular orbit occurs when $\dot{r}=0$ and $dV_{\text{eff}}/dr=0$. By solving these conditions, one obtains the specific energy $\tilde{E}$ and angular momentum $\tilde{L}$, given by
\begin{eqnarray}
\tilde{E} = \frac{-g_{tt} + g_{t\phi}\Omega_\phi}{\sqrt{ -g_{tt} - 2g_{t\phi}\Omega_\phi - g_{\phi\phi}\Omega_\phi^2 }}, \label{a9} \\
\tilde{L} = \frac{g_{t\phi} + g_{\phi\phi}\Omega_\phi}{\sqrt{ -g_{tt} - 2g_{t\phi}\Omega_\phi - g_{\phi\phi}\Omega_\phi^2 }}.\label{a10}
\end{eqnarray}
In these expressions, $\Omega_{\phi}$  represents the angular velocity of the particles in circular orbits computed as
\begin{equation}
\Omega_\phi = \frac{ -\partial_r g_{t\phi} \pm \sqrt{ (\partial_r g_{t\phi})^2 - (\partial_r g_{tt})(\partial_r g_{\phi\phi}) } }{ \partial_r g_{\phi\phi} }.\label{a11}
\end{equation}
Here, `$+$' signifies co-rotating while `$-$' corresponds to counter-rotating orbits. For the case of co-rotating orbits, the angular momentum is directed along the BH spin, whereas for counter-rotating orbits, it is antiparallel to the direction of the BH spin.

In this study, we explore QPOs by relating them to the orbital frequency $\nu_\phi$, the radial epicyclic frequency $\nu_r$, and the vertical epicyclic frequency $\nu_{\theta}$, which correspond to circular orbits. The orbital frequency, also known as the Keplerian frequency, is expressed as
\begin{eqnarray}
    \nu_\phi=\frac{\Omega_\phi}{2\pi}.
\end{eqnarray}
The vertical and radial epicyclic frequencies are computed by assuming small perturbations near the circular equatorial orbit, and the motion of the particle is expressed as
\begin{eqnarray}
\theta(t) = \frac{\pi}{2} + \delta \theta(t), \quad r(t) = r_0 + \delta r(t), \label{a14}
\end{eqnarray}
here, $\delta r(t)$ and $\delta \theta(t)$ represent the small perturbations governing the following equations.
\begin{eqnarray}
\frac{d^2 \delta \theta(t)}{dt^2} + \Omega_\theta^2 \delta \theta(t) = 0,\label{a15}\\
\frac{d^2 \delta r(t)}{dt^2} + \Omega_r^2 \delta r(t) = 0, \label{a16}
\end{eqnarray}
where
\begin{eqnarray}
\Omega_\theta^2 = -\frac{1}{2 g_{\theta\theta} \dot{t}^2} \left. \frac{\partial^2 V_{\text{eff}}}{\partial \theta^2} \right|_{\theta = \frac{\pi}{2}}, \label{a17}\\
\Omega_r^2 = -\frac{1}{2 g_{rr} \dot{t}^2} \left. \frac{\partial^2 V_{\text{eff}}}{\partial r^2} \right|_{\theta = \frac{\pi}{2}}, \label{a18}
\end{eqnarray}
The vertical and radial frequencies epicyclic are obtained by using Eqs.~(\ref{a17}) and (\ref{a18})
\begin{eqnarray}
\nu_\theta &=& \frac{\Omega_\theta}{2\pi},\\
\nu_r &=& \frac{\Omega_r}{2\pi}. 
\end{eqnarray}
Appendix A contains the exact expressions of $\nu_\phi$, $\nu_r$, and $\nu_\theta$ for the accelerating BH. When examining equatorial circular orbits for a test particle, the radial oscillations relative to the mean orbit are characterized by the radial epicyclic frequency $\nu_{r}$, and the oscillations perpendicular to the equatorial plane are characterized by the vertical epicyclic frequency $\nu_{\theta}$. 

\section{The frequency prescriptions for quasiperiodic oscillations}

Numerous theoretical models have been proposed to explain the QPO phenomenon. In this section, we consider three typical QPO models, namely, the RP model (linking QPOs to orbital, radial, and vertical epicyclic frequencies), the PR model (describing nonlinear coupling and parametric resonances between radial and vertical oscillations), and the FR model (where resonant response is driven by disk or external perturbations). 

\subsection{Relativistic Precession Model}

The RP model is used to study high-frequency quasi-periodic oscillations (HFQPOs) in neutron star sources and has also been applied to HFQPOs observed in BHs \cite{Stella:1999sj}. For this model, the frequency of periastron precession, $\nu_{\rm per}$, and the frequency of nodal precession, $\nu_{\rm nod}$, are defined as
\begin{eqnarray}
\nu_{\rm per} &=&  \nu_\phi - \nu_r, \\
\nu_{\rm nod} &=&  \nu_\phi - \nu_\theta.
\end{eqnarray}
According to the RPM for the X-ray BH binaries~\cite{Stella:1997tc, Stella:1998mq, Stella:1999sj}, the following three frequencies, $\nu_\phi$, $\nu_{\rm per}$, and $\nu_{\rm nod}$, correspond to the observed upper high-frequency QPO ($\nu_U$), lower high-frequency QPO ($\nu_L$), and low-frequency type-C QPO ($\nu_C$)
\begin{eqnarray}
\nu_U = \nu_\phi,\;\; \nu_L = \nu_{\rm per}, \;\;\nu_C = \nu_{\rm nod}. \label{RP}
\end{eqnarray}
\subsection{Parametric Resonance Model}
The persistent detection of a 3:2 ratio in twin-peak high-frequency QPOs from neutron star and BH systems suggests that these oscillations originate from resonances between different accretion disk motion modes \cite{Kluzniak:2002bb, Abramowicz:2001bi, Abramowicz:2003xy, Rebusco:2004ba,  Abramowicz:2001bi, Abramowicz:2004je}. In this formulation, small perturbations in the vertical and radial directions near the equatorial geodesics are regarded as distinct harmonic oscillations, which can be identified by the vertical ($\nu_{\theta}$) and radial ($\nu_{r}$) epicyclic frequencies, respectively. According to the PR model, radial oscillations are more pronounced than vertical oscillations in thin accretion disks ($\delta r > \delta \theta$). They can parametrically produce vertical oscillations when the resonance condition $\nu_{r}/\nu_{\theta} = 2/n$ holds, where $n$ is a positive integer. For rotating BHs, where $\nu_{\theta} > \nu_{r}$ often holds, the resonance is most prominent for $n = 3$, which obviously leads to the usual 3:2 frequency ratio. For this model, the lower and upper frequencies are 
\begin{eqnarray}
\nu_L = \nu_{r} ,\;\; \nu_U = \nu_\theta \label{PR}
\end{eqnarray}
\subsection {Forced Resonance Model}
Accretion flows are often not adequately described by the thin Keplerian disk \cite{Kluzniak:2002bb, Abramowicz:2001bi, 2001AcPPB..32.3605K, 2005A&A...436....1T} due to the influence of pressure, viscosity, or magnetic stresses within the accretion flow. This leads to a non-linear relation between $\delta r$ and $\delta \theta$, along with the previously mentioned parametric resonance. Numerical simulations have verified that a resonance of vertical oscillations induced by radial oscillations can occur through pressure coupling~\cite{Abramowicz:2001bi, Lee:2004bp}.
These nonlinear couplings between $\delta r$ and $\delta\theta$ are often described using a mathematical ansatz.
\begin{eqnarray}
{\delta \ddot\theta} + \omega_{\theta}^{2}\delta \theta = -\omega_{\theta}^{2}\,\delta r\,\delta \theta + \mathcal{F}_{\theta}(\delta \theta) \label{fmodel}
\end{eqnarray}
where $\delta r=Acos(\omega_{r}t)$ and $\mathcal{F}_{\theta}$ signifies the non-linear terms in $\delta \theta$. By solving Eq. \ref{fmodel} one can obtain
\begin{equation}
\frac{\nu_{\theta}}{\nu_{r}} = \frac{m}{n}, \qquad \text{where $m$ and $n$ are natural numbers.}
\end{equation}
For the case of forced resonance mode $m:n=3:1$ the upper and lower frequencies are given by 
\begin{eqnarray}
\nu_{U} &=&  \nu_\theta\\
\nu_{L} &=&  \nu_\theta - \nu_r. \label{FR}
\end{eqnarray}

\section{Observational Analysis}

\begin{table*}[t]
\begin{ruledtabular}
\caption{The QPOs from the X-ray binaries that have been selected for investigation, including their mass, orbital frequencies, periastron precession frequencies, and nodal precession frequencies.}
\label{tab: I}
\begin{tabular}{c|c|c|c|c|c|c|c}
& GRO J1655--40 & XTE J1550--564 & XTE J1859+226 & GRS 1915+105 & H1743--322 & M82\,X{-1} & Sgr A$^{*}$ \\
\hline
$M~(M_{\odot})$
& $5.4 \pm 0.3$~\cite{Motta:2013wga}
& $9.1 \pm 0.61$~\cite{Remillard:2002cy,Orosz:2011ki}
& $7.85 \pm 0.46$~\cite{Motta:2022rku}
& $12.4^{+2.0}_{-1.8}$~\cite{Remillard:2006fc}
& $\gtrsim 9.29$~\cite{Ingram:2014ara}
& $415 \pm 63$~\cite{Pasham2014}
& $(3.5\text{--}4.9)\cdot 10^{6}$~\cite{Ghez:2008ms, Gillessen:2008qv}
\\
$\nu_{U}\,(\mathrm{Hz})$
& $441 \pm 2$~\cite{Motta:2013wga}
& $276 \pm 3$~\cite{Remillard:2002cy}
& $227.5^{+2.1}_{-2.4}$~\cite{Motta:2022rku}
& $168 \pm 3$~\cite{Remillard:2006fc}
& $240 \pm 3$~\cite{Ingram:2014ara}
& $5.07 \pm 0.06$~\cite{Pasham2014}
& $(1.45 \pm 0.16)\times 10^{-3}$~\cite{Stuchlik:2008fy}
\\
$\nu_{L}\,(\mathrm{Hz})$
& $298 \pm 4$~\cite{Motta:2013wga}
& $184 \pm 5$~\cite{Remillard:2002cy}
& $128.6^{+1.6}_{-1.8}$~\cite{Motta:2022rku}
& $113 \pm 5$~\cite{Remillard:2006fc}
& $165^{+9}_{-5}$~\cite{Ingram:2014ara}
& $3.32 \pm 0.06$~\cite{Pasham2014}
& $(0.89 \pm 0.04)\times 10^{-3}$~\cite{Stuchlik:2008fy}
\\
$\nu_{C}\,(\mathrm{Hz})$
& $17.3 \pm 0.1$~\cite{Motta:2013wga}
& --
& $3.65 \pm 0.01$~\cite{Motta:2022rku}
& --
& $9.44 \pm 0.02$~\cite{Ingram:2014ara}
& --
& --
\end{tabular}
\end{ruledtabular}
\end{table*}

\begin{table}
\centering
\begin{ruledtabular}
\caption{The prior range on the model parameters. We choose a uniform \((\mathcal{U})\) and a Gaussian range \((\mathcal{N(\mu, \sigma)})\) for the selected parameters for entire observational data.}
\label{tab:placeholder}
\begin{tabular}{cc}
Parameters & Prior Range\\
\hline
\(m(m_\odot)\)& $ \mathcal{U} [1,10^8]$\\  
\(a/M\) & \( \mathcal{N}(0.4, 0.05)\) \\
\(r/M\) & \(\mathcal{N} (5.5, 0.5)\) \\
\(A \cdot m\) \text{for Sgr A} & $ \mathcal{U} [0,1.5]$
\end{tabular}
\end{ruledtabular}
\end{table}

In this section, we discuss the observational data used to constrain the dimensionless parameters of the current BH model, namely its mass $(m)$, spin parameter $(a/m)$, orbital radius parameter $(r/m)$, and acceleration parameter $(A \cdot m)$. The observational data corresponding to distinct X-ray timing sources are summarized in Table~\ref{tab: I}. We consider a total of seven independent observational samples, where the corresponding BH mass (in solar mass units), as well as the upper, lower, and centroid frequencies, are listed. In the table, missing frequency measurements are denoted by a dash. 
		
		To constrain the model parameters, we adopt three different theoretical frameworks: the Relativistic Precession (RP), Parametric Resonance (PR), and Forced Resonance (FR) models. For each of these, we employ the observational data corresponding to the triplet $\{\nu_{U}, \nu_{L}, \nu_{C}\}$. The posterior distribution of the model parameters is then computed using Bayes' theorem:
		\begin{equation}
			P(\boldsymbol{\theta} | D, H) = \frac{P(D | \boldsymbol{\theta}, H) \, P(\boldsymbol{\theta} | H)}{P(D | H)} \, ,
		\end{equation}
		where $\boldsymbol{\theta}$, $D$, and $H$ denote the parameter vector, the data vector, and the model hypothesis, respectively. The left-hand side represents the posterior probability of the parameters given the data, while $\mathcal{L} \equiv P(D | \boldsymbol{\theta}, H)$ is the likelihood, defined as
		\begin{equation}
			\mathcal{L} = \exp\left(-\frac{1}{2} \chi^2 \right) \, ,
		\end{equation}
		with the chi-squared quantity $\chi^2$ given by
		\begin{equation}
			\chi^2 = \sum_{i=1}^{N} \left( \frac{D_{i, \mathrm{Obs}} - D_{i, \mathrm{Model}}}{\sigma_i} \right)^2 \, .
		\end{equation}
		Here, $\sigma_i$ denotes the statistical uncertainty associated with each observational measurement. The prior distribution $P(\boldsymbol{\theta} | H)$ encodes our assumptions about the parameters before considering the data. In this work, we adopt uniform (flat) priors for parameters with well-bounded domains and broad Gaussian priors for parameters with uncertain but approximately known ranges \cite{Padilla:2019mgi}.
		
		The likelihood evaluation is implemented in a Python-based pipeline developed for the current BH model. For posterior sampling, we employ the \emph{dynamic nested sampling} algorithm \texttt{dynesty}, which is particularly efficient for multimodal or degenerate posteriors \cite{Higson:2018cwj}. The resultant posterior samples are analyzed using the \texttt{GetDist} package to extract marginalized constraints on each parameter and to generate one- and two-dimensional posterior distributions \cite{Lewis:2019xzd}. Parameter estimates are quoted at the 68\% confidence level (CL) unless otherwise specified.
		
\begin{figure*}
\centering
\includegraphics[scale=0.29]{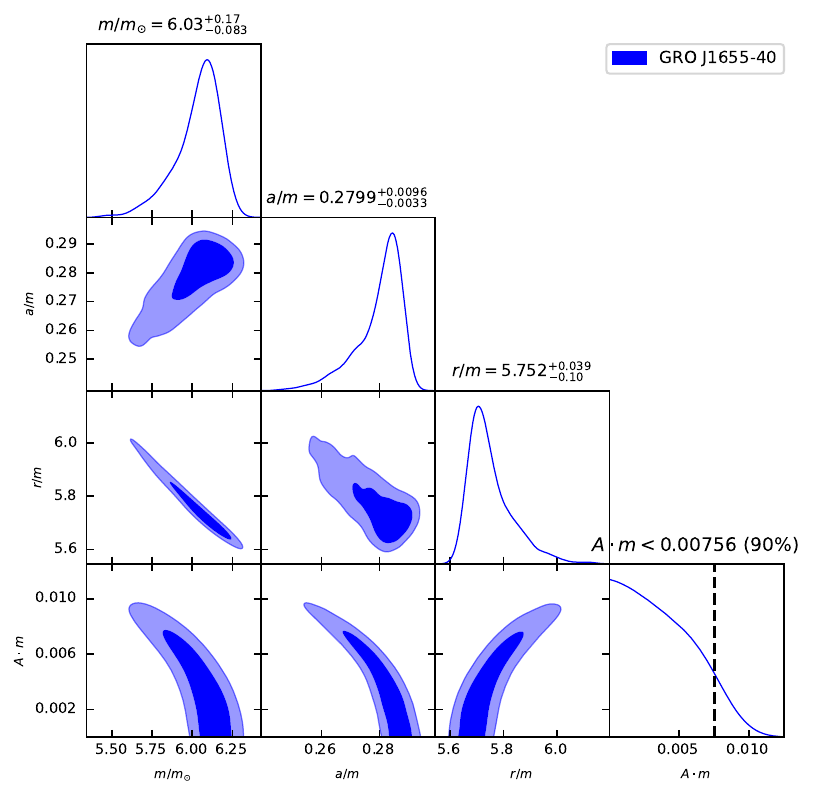}
\includegraphics[scale=0.29]{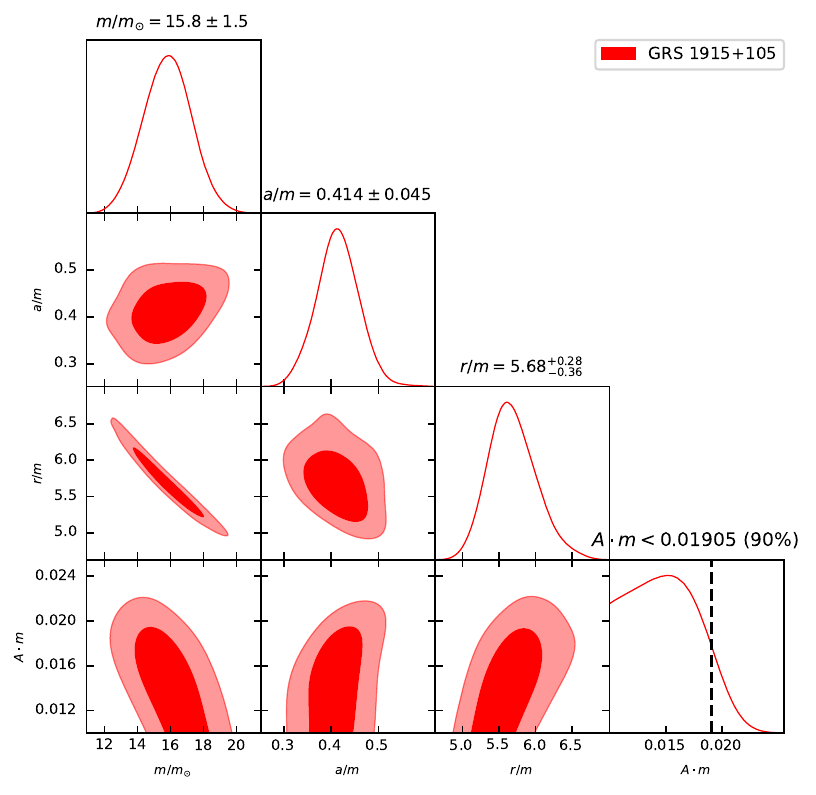}
\includegraphics[scale=0.29]{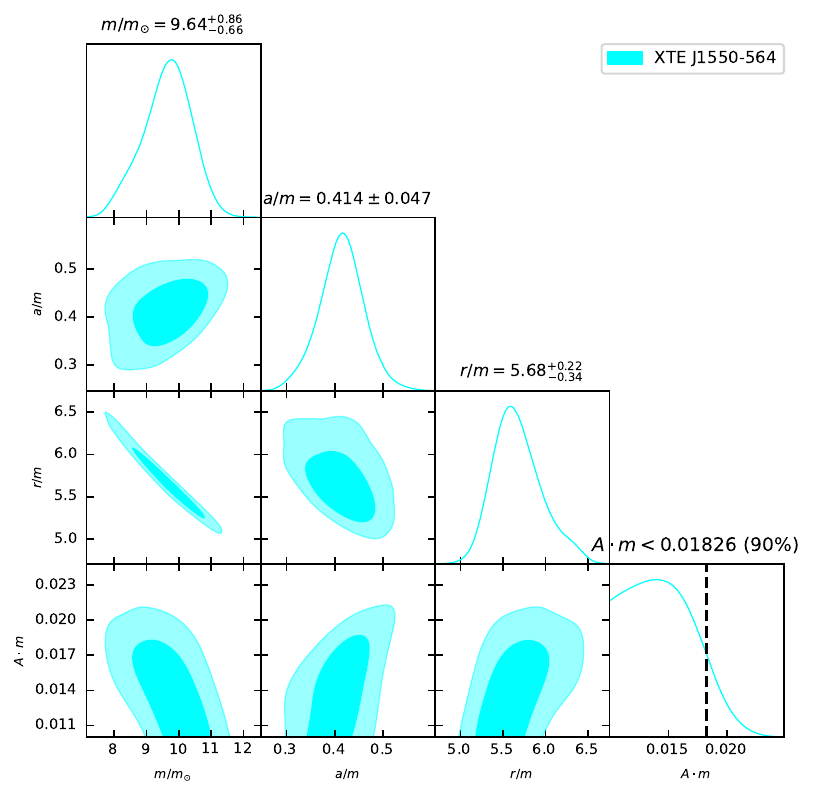}
\includegraphics[scale=0.29]{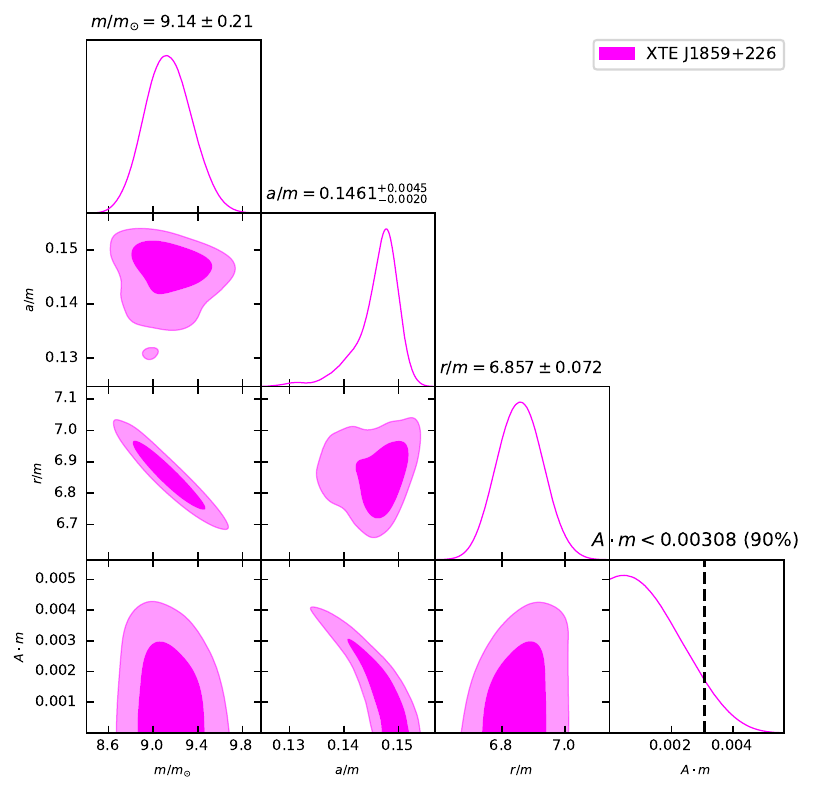}
\includegraphics[scale=0.29]{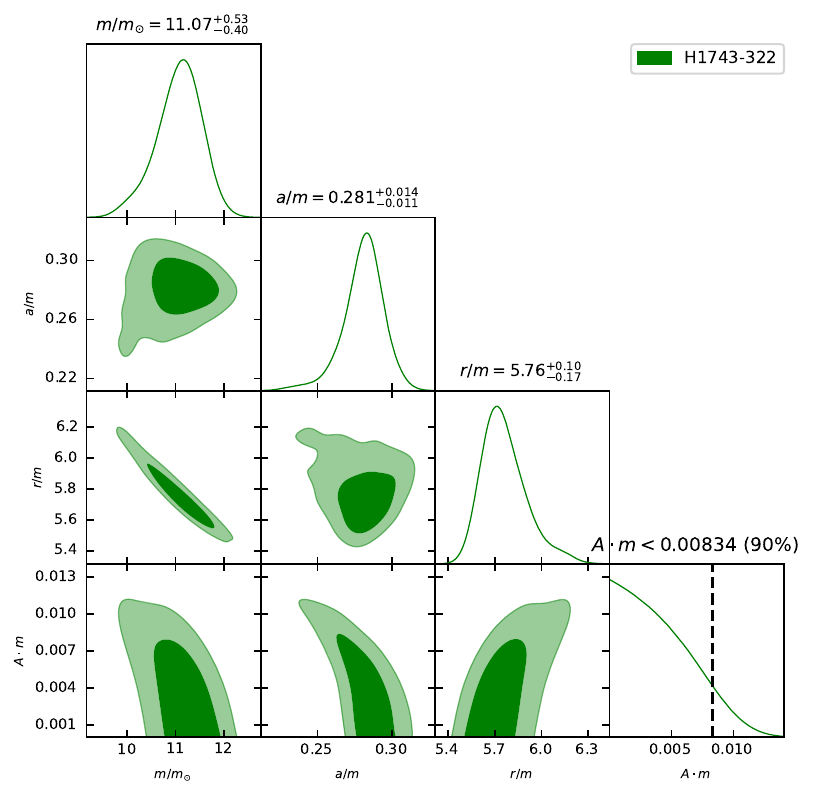}
\includegraphics[scale=0.29]{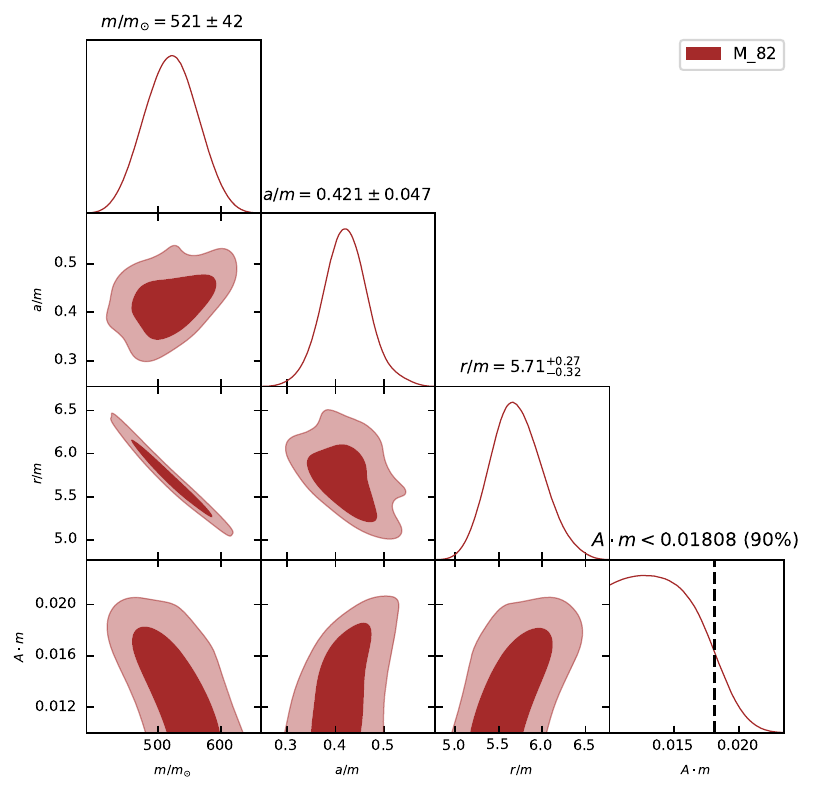}
\includegraphics[scale=0.29]{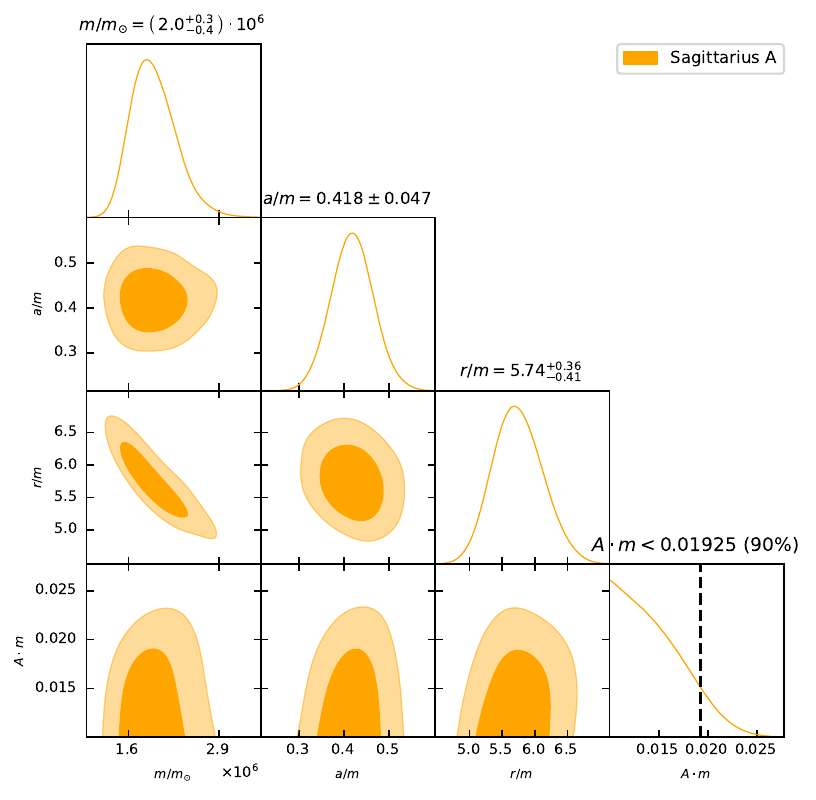}
\caption{The posterior distributions of the BH mass \(M\), spin parameter \(a/M\), orbital radius \(r/M\), and dimensionless acceleration parameter \(m\cdot A\), obtained within the RP model using the observed QPOs of X-ray binaries, are given in Table~\ref{tab: I}. The corner plots display the marginalized posterior distributions, with the shaded regions corresponding to the 68\% and 90\% confidence intervals for each source.
}
\label{1a}
\end{figure*}

		In this analysis, we impose uniform priors on the BH mass and acceleration parameter, while Gaussian priors with large dispersions are applied to the spin and orbital radius parameters. The ranges of all priors are listed in Table~\ref{tab:placeholder}. The resulting one- and two-dimensional posterior distributions for each model are shown in Figs.~\ref{1a}, \ref{2a}, and \ref{3a}, where the central parameter values correspond to the 68\% CL. For the acceleration parameter, we report an upper bound at the 90\% CL by restricting the posterior samples to the physically motivated range $A \cdot m > 0$. In most cases, the posterior probability density peaks near zero, reflecting the limited precision of the current data, which does not yet allow a statistically significant deviation from the Kerr BH solution. Consequently, quoting an upper bound on the acceleration parameter captures the essential physical implications of the present analysis.

\section{Results and Discussion}

\subsection{Case I: Relativistic Precession Model (RP)}
	
The central values of the model parameters at the 68\% confidence level (CL) are listed in Table~\ref{tab:bestfit} for each model. In the case of the RP model, the inferred BH mass is found to be consistent with the corresponding observational estimates. A mild shift in the best-fit mass values is observed relative to the observationally inferred masses listed in Table~\ref{tab: I}. 
	
The spin parameter $(a/m)$ varies in the range \(0.14 \lesssim a/m \lesssim 0.43\) across different sources, while the orbital radius parameter $(r/m)$ lies within the interval \(5.6 \lesssim r/m \lesssim 6.9\). These values remain broadly consistent across all observational samples. On the other hand, the 90\% CL upper bound on the acceleration parameter \((A \cdot m)\) is found to be in the range \(0.003\text{--}0.02\).  
	
Overall, the results indicate that the frame-dragging effect is primarily governed by the spin parameter $(a/m)$, with the acceleration parameter contributing only a small perturbative effect. Consequently, the spacetime remains effectively Kerr-like, as the inferred acceleration leads to only negligible deviations from the Kerr geometry.

\subsection{Case II: Parametric Resonance Model (PR)}

\begin{figure*}
\centering
\includegraphics[scale=0.29]{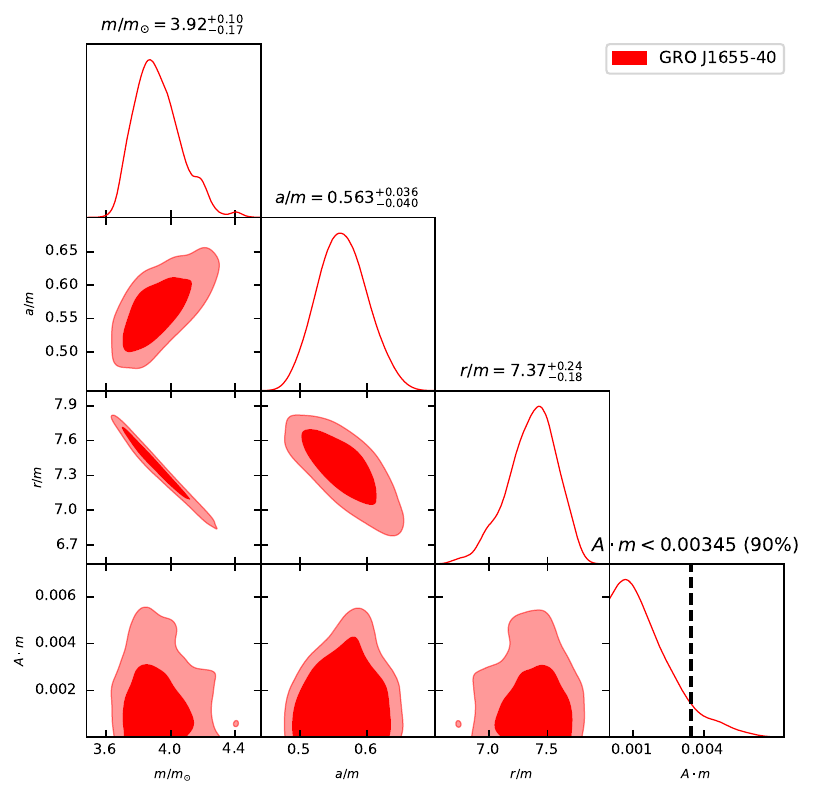}
\includegraphics[scale=0.29]{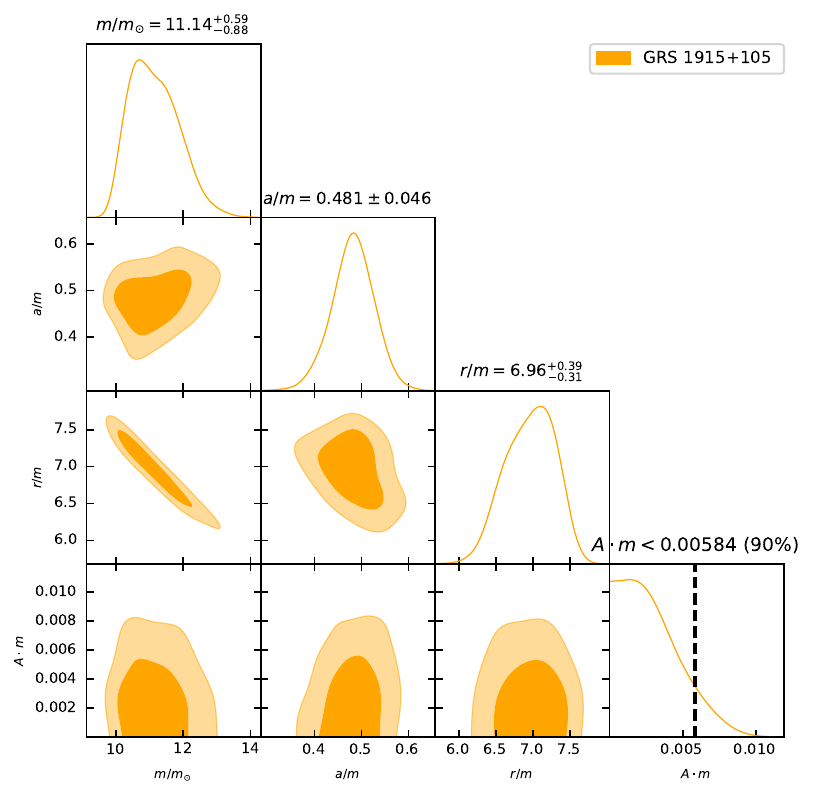}
\includegraphics[scale=0.29]{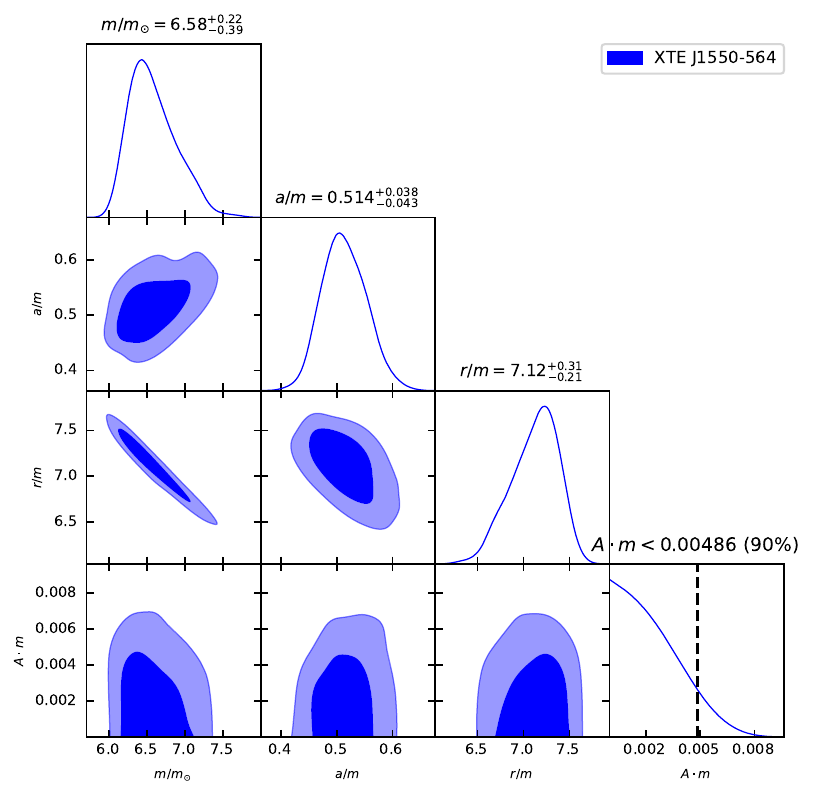}
\includegraphics[scale=0.29]{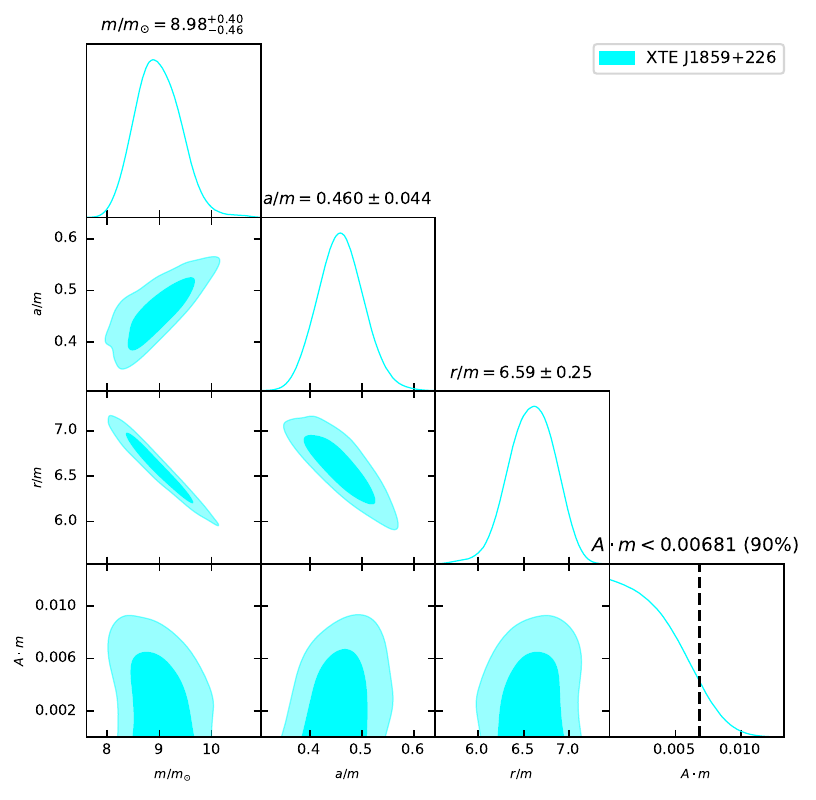}
\includegraphics[scale=0.29]{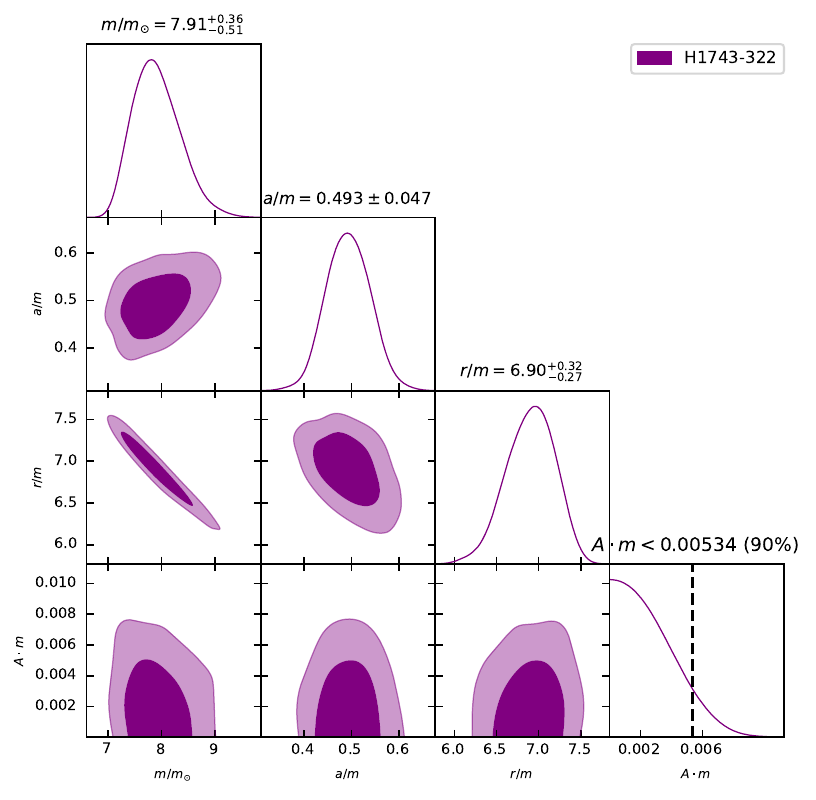}
\includegraphics[scale=0.29]{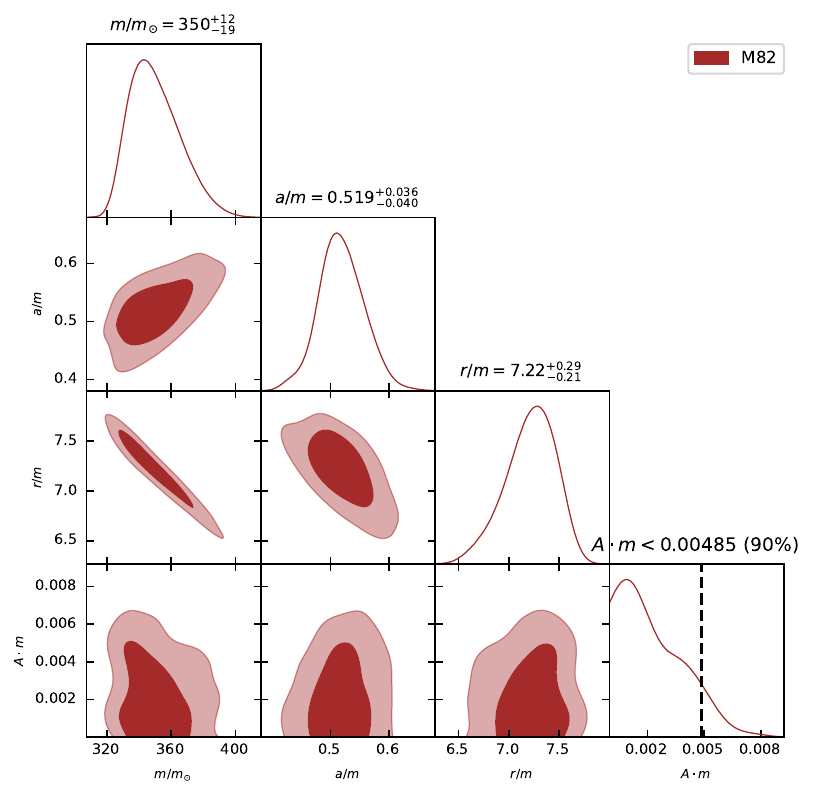}
\includegraphics[scale=0.29]{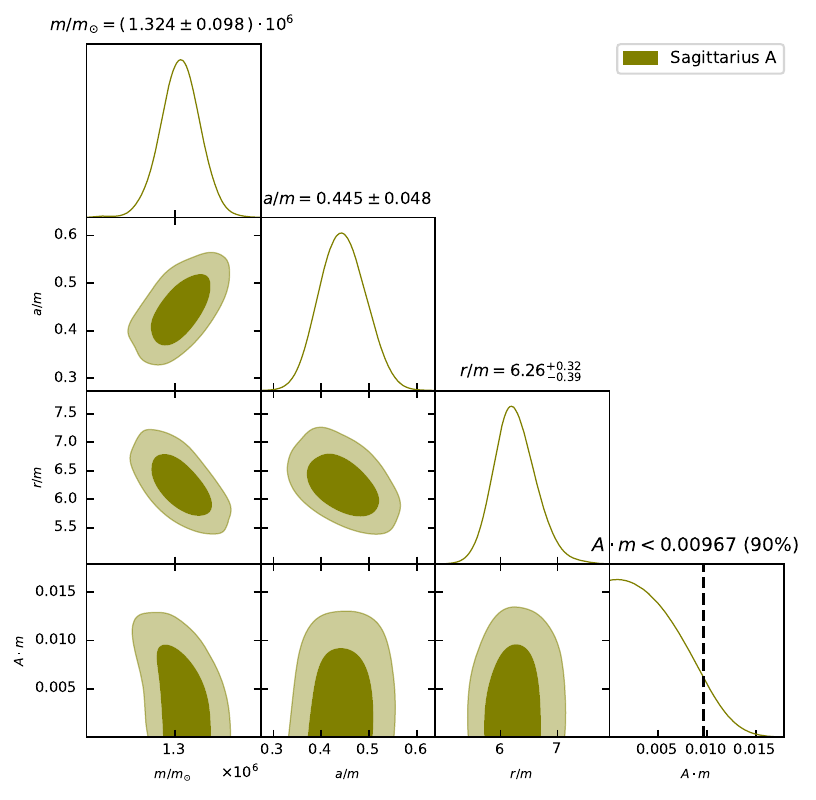}
\caption{The posterior distributions of the BH mass \(M\), spin parameter \(a/M\), orbital radius \(r/M\), and dimensionless accelerating parameter \(m\cdot A\), obtained within the PR model using the observed QPOs of X-ray binaries, are given in Table~\ref{tab: I}. The corner plots represent the marginalized posterior distributions, with the shaded regions corresponding to the 68\% and 90\% confidence intervals for each source.}
\label{2a}
\end{figure*}

The posterior distributions for this model are shown in Fig.~\ref{2a}, and the corresponding best-fit values are listed in Table~\ref{tab:bestfit}. In this case, the inferred BH masses fall within ranges consistent with the observational estimates reported in Table~\ref{tab: I}, with only mild deviations. The spin parameters for all sources are nearly consistent, lying in the range \(0.44 \lesssim a/m \lesssim 0.57\). A similar trend is observed for the orbital radius parameter, which spans \(6.1 \lesssim r/m \lesssim 7.4\). Compared to the RP model, both the spin and orbital radius parameters take marginally higher values.
		
From the posterior distributions, we find that the mass parameter is strongly anti-correlated with the orbital radius: as the BH mass decreases, the corresponding orbital radius increases. The spin parameter shows a mild positive correlation with the mass parameter, indicating that higher-mass configurations tend to exhibit slightly larger spins. Similarly, the spin and orbital radius parameters are weakly anti-correlated. These trends are qualitatively consistent with those obtained for the RP model.
		
The upper bounds on the acceleration parameter $(A \cdot m)$ are consistent across all observational samples. The posterior analysis further reveals that this parameter exhibits little to no correlation with the other quantities in the PR model, in contrast to the RP model, where $A \cdot m$ is negatively correlated with the mass parameter and positively correlated with the orbital radius. For the RP model, the acceleration parameter was found to be strongly anti-correlated with the spin for GRO~J1655--40, XTE~J1859+226, and H1743--322, while it was positively correlated for the remaining sources.
		
Furthermore, the PR model generally favors higher spin values and slightly lower masses than those obtained from the RP model. This behavior arises because the resonance condition constrains the frequency ratio independently of geometric corrections, requiring a stronger frame-dragging effect (larger $a/m$) to reproduce the observed QPO pairings. Our analysis indicates that the dimensionless acceleration parameter $(A \cdot m)$ has a negligible impact on the disk dynamics in the resonance regime. The consistently small inferred values of $(A \cdot m)$ across all sources suggest that any plausible acceleration of the central compact object is too weak to affect the observable timing signals within current sensitivity limits. Consequently, the parameter $A \cdot m$ introduces only a minor perturbation to the resonant oscillation structure, implying that the spacetime around the accelerating BH remains effectively Kerr-like for all the analyzed sources.

\subsection{Case III: Forced Resonance Model (FR)}

\begin{figure*}
\centering
\includegraphics[scale=0.29]{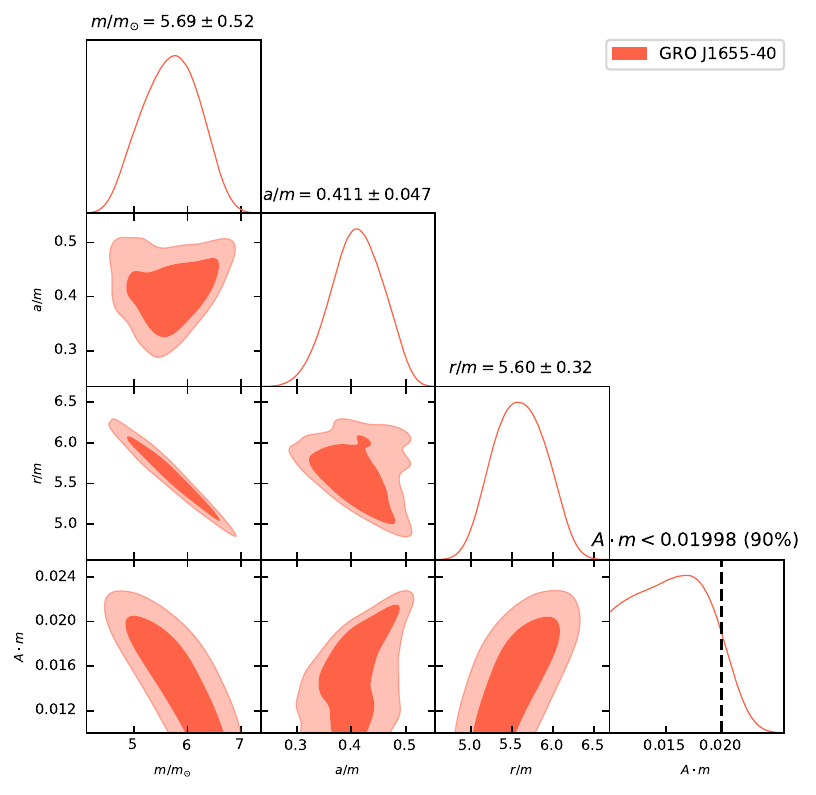}
\includegraphics[scale=0.29]{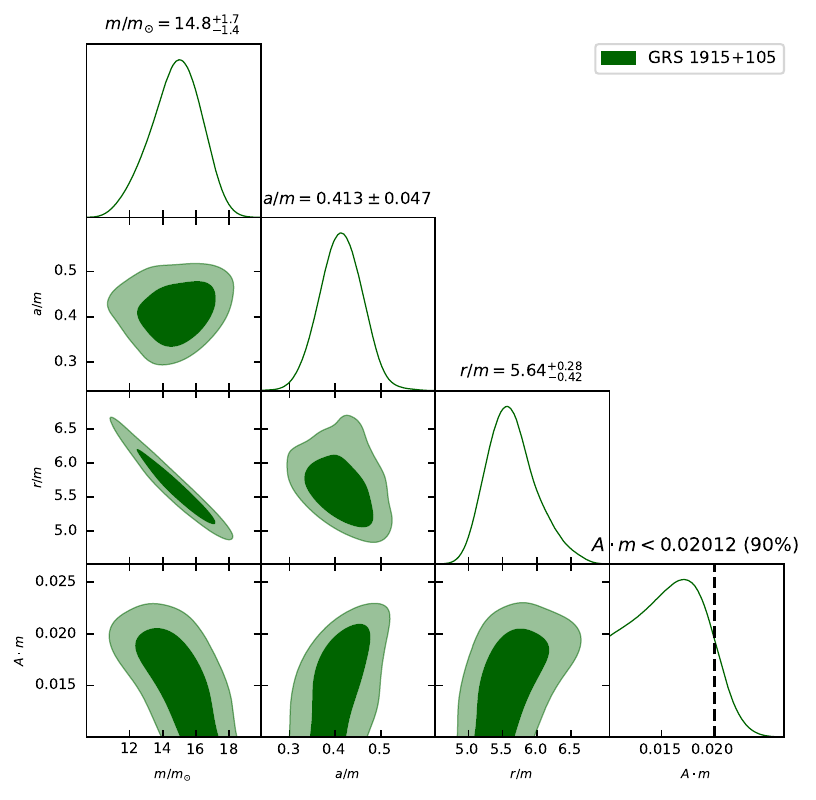}
\includegraphics[scale=0.29]{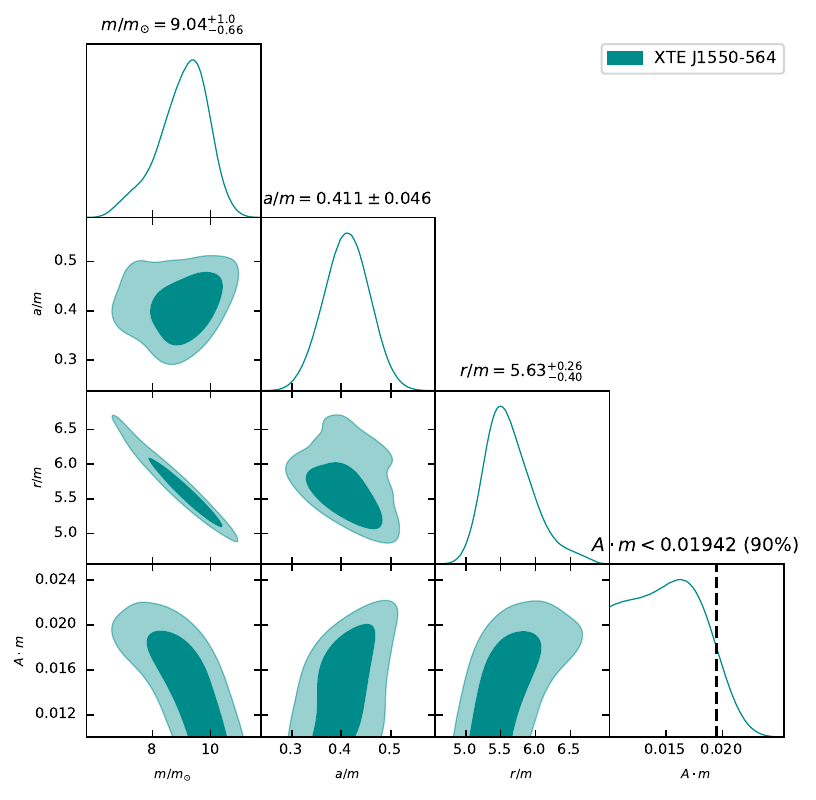}
\includegraphics[scale=0.29]{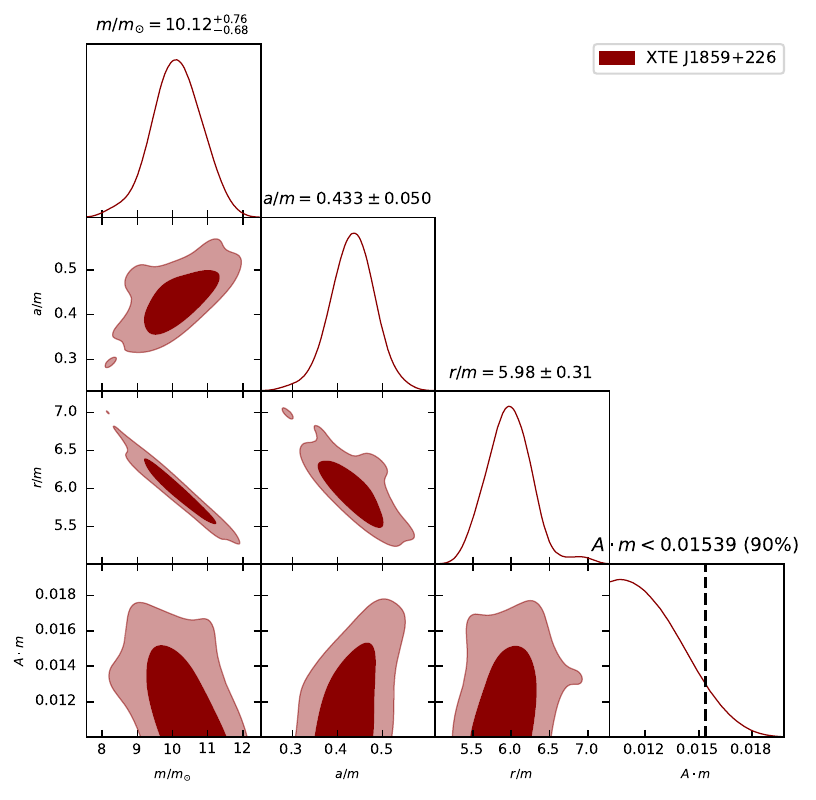}
\includegraphics[scale=0.29]{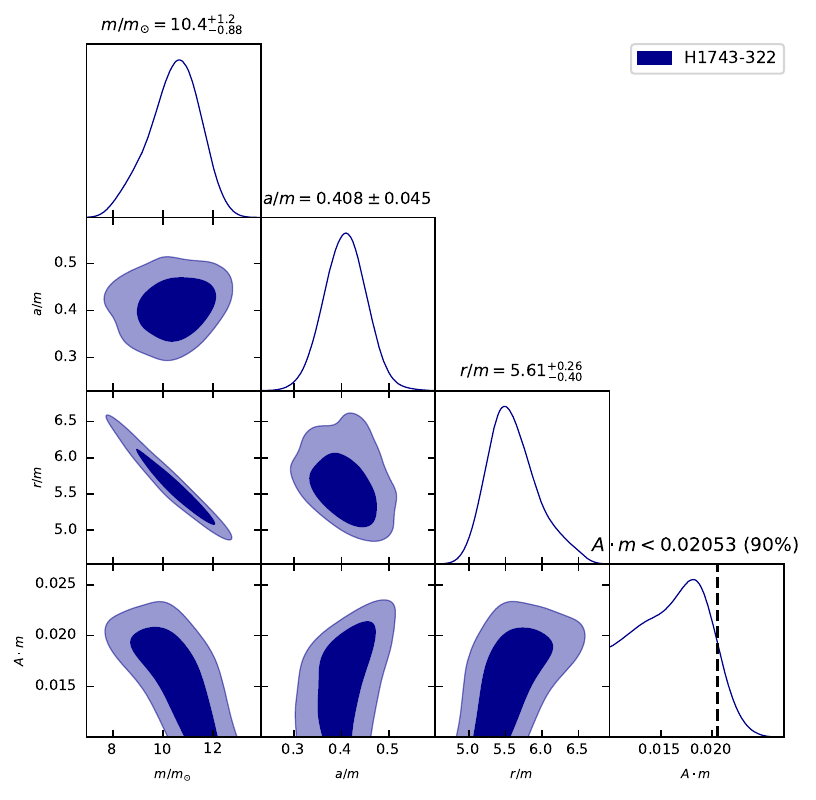}
\includegraphics[scale=0.29]{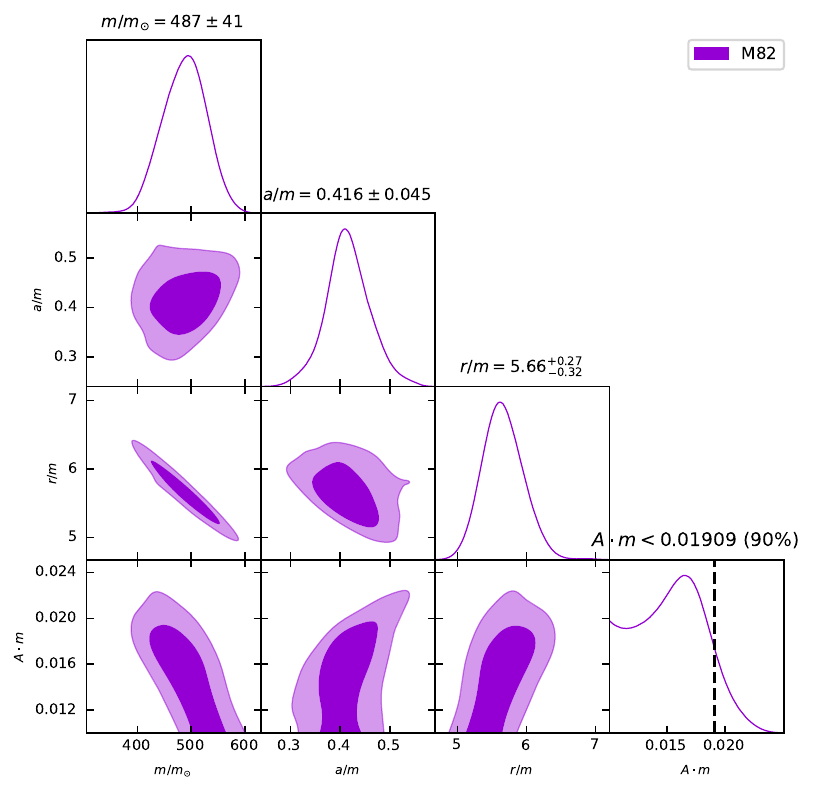}
\includegraphics[scale=0.29]{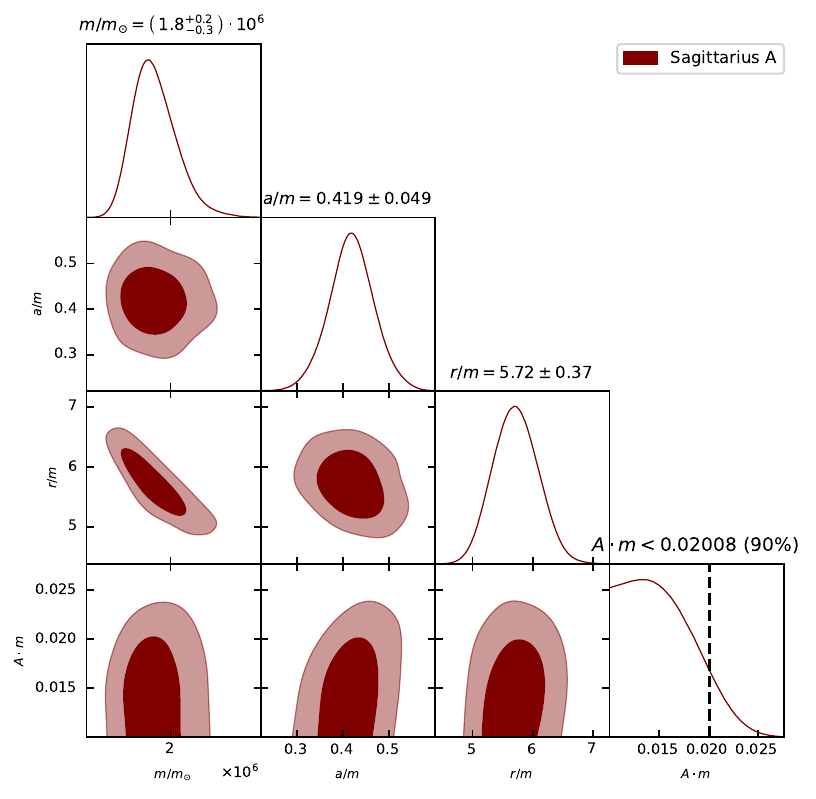}
\caption{The posterior distributions of the BH mass $M$, spin parameter $a/M$, orbital radius $r/M$, and dimensionless acceleration parameter $m\cdot A$, obtained within the forced resonance model using the observed QPOs of X-ray binaries, are provided in Table~\ref{tab: I}. The corner plots display the marginalized posterior distributions, with the shaded regions corresponding to the 68\% and 90\% confidence intervals for each source.}
\label{3a}
\end{figure*}

For the FR model, the posterior distributions are shown in Fig.~\ref{3a}, where the inferred mass parameters, listed in Table~\ref{tab:bestfit}, are found to be in close agreement with the observed values. Unlike the previous two cases, this model yields highly consistent spin and orbital radius parameters across all observational samples, indicating a stronger overall agreement with the data. The mass parameter exhibits a negative correlation with both the orbital radius and the acceleration parameter, while showing a mild positive correlation with the spin parameter—consistent with the trend observed in the RP model. 
	
The acceleration parameter in this case is anti-correlated with the mass parameter and positively correlated with both the spin and orbital radius parameters, showing a slightly different correlation pattern from that of the RP model. The corresponding 90\% CL upper limits on the acceleration parameter $(A \cdot m)$ are remarkably consistent across the samples, typically lying in the range \(0.019 \lesssim A \cdot m \lesssim 0.02\), except for XTE~J1859+226, where the upper limit is slightly lower, \(A \cdot m < 0.0154\).
	
The relatively higher upper bound obtained for this model compared to the PR case suggests that, with improved observational precision, the nonlinear coupling between the radial and vertical oscillation modes ($\delta r$ and $\delta \theta$) can be further tested. Such effects could influence the rotational dynamics of the accretion flow and may become sensitive to the acceleration of the central compact object.

\begin{table*}
\begin{ruledtabular}
\caption{The constraint on BH parameters at 68\% confidence level for distinct frequency models. The upper bound on \(m \cdot A \) is obtained at \(90\%\) confidence level.}
\label{tab:bestfit}
\begin{tabular}{ccccc}
Model & $m/(m_{\odot})$ & $a/m$ & $r/m$ & $m \cdot A \ (90\%)$ \\
\hline
\multicolumn{5}{c}{\textbf{Relativistic Precession Model}} \\
\hline
GRO J1655--40   & $6.03^{+0.17}_{-0.083}$ & $0.2799^{+0.0096}_{-0.0033}$ & $5.752^{+0.039}_{-0.10}$ & $<0.00756$ \\
GRS 1915+105    & $15.8 \pm 1.5$ & $0.414 \pm 0.045$ & $5.68^{+0.28}_{-0.36}$ & $<0.01905$ \\
XTE J1550--564  & $9.64^{+0.86}_{-0.66}$ & $0.414 \pm 0.047$ & $5.68^{+0.22}_{-0.34}$ & $<0.0182$ \\
XTE J1859+226   & $9.14 \pm 0.21$ & $0.1461^{+0.0045}_{-0.0020}$ & $6.857 \pm 0.072$ & $<0.00308$ \\
H1743--322      & $11.07^{+0.53}_{-0.40}$ & $0.281^{+0.014}_{-0.011}$ & $5.76^{+0.10}_{-0.17}$ & $<0.00834$ \\
M82        & $521 \pm 42$ & $0.421 \pm 0.047$ & $5.71^{+0.27}_{-0.32}$ & $<0.01808$ \\
Sgr A*    & $(2.0^{+0.3}_{-0.4})\times 10^{6}$ & $0.418 \pm 0.047$ & $5.74^{+0.36}_{-0.41}$ & $<0.01925$ \\
\hline
\multicolumn{5}{c}{\textbf{Parametric Resonance Model}} \\
\hline
GRO J1655--40   & $3.92^{+0.10}_{-0.17}$ & $0.563^{+0.036}_{-0.040}$ & $7.37^{+0.24}_{-0.18}$ & $<0.00345$ \\
GRS 1915+105    & $11.4_{-0.88}^{+0.59}$ & $0.481 \pm 0.046$ & $6.96^{+0.39}_{-0.31}$ & $<0.00584$ \\
XTE J1550--564  & $6.58^{+0.22}_{-0.39}$ & $0.514_{0.043}^{0.038}$ & $7.12^{+0.31}_{-0.21}$ & $<0.00486$ \\
XTE J1859+226   & $8.98_{-0.46}^{0.40}$ & $0.46\pm 0.044$ & $6.59 \pm 0.25$ & $<0.00681$ \\
H1743--322      & $7.91^{+0.36}_{-0.51}$ & $0.493\pm 0.047$ & $6.90^{+0.32}_{-0.27}$ & $<0.00534$ \\
M82     & $350_{-19}^{+12}$ & $0.421 \pm 0.519_{-0.040}^{+0.036}$ & $7.22^{+0.29}_{-0.21}$ & $<0.00485$ \\
Sgr A*  & $(1.324\pm 0.098)\times 10^{6}$ & $0.445 \pm 0.048$ & $6.26^{+0.32}_{-0.39}$ & $<0.00967$ \\
\hline
\multicolumn{5}{c}{\textbf{Forced Resonance Model}} \\
\hline
GRO J1655--40   & $5.69\pm 0.52$ & $0.411\pm0.047$ & $5.60\pm0.32$ & $<0.01998$ \\
GRS 1915+105    & $14.8_{-1.4}^{+1.7}$ & $0.413 \pm 0.047$ & $5.64^{+0.28}_{-0.42}$ & $<0.02012$ \\
XTE J1550--564  & $9.04^{+0.1}_{-0.66}$ & $0.411 \pm 0.046$ & $5.63^{+0.26}_{-0.40}$ & $<0.01942$ \\
XTE J1859+226   & $10.12_{0.68}^{0.76}$ & $0.433\pm 0.050$ & $5.93 \pm 0.31$ & $<0.01539$ \\
H1743--322      & $10.4^{+1.2}_{-0.88}$ & $0.408\pm0.045$ & $5.61^{+0.26}_{-0.40}$ & $<0.02053$ \\
M82   & $487 \pm 41$ & $0.416 \pm 0.045$ & $5.66^{+0.27}_{-0.32}$ & $<0.01909$ \\
Sgr A*  & $(1.8^{+0.2}_{-0.3})\times 10^{6}$ & $0.419 \pm 0.049$ & $5.72\pm 0.37$ & $<0.02008$ 
\end{tabular}
\end{ruledtabular}
\end{table*}

\section{Conclusions}
\renewcommand{\theequation}{4.\arabic{equation}} \setcounter{equation}{0}

This article focuses on the investigation of QPOs observed in X-ray binaries in the vicinity of an accreting BH. For this purpose, we computed the fundamental frequencies resulting from the motion of a test particle around the accelerating Kerr spacetime. By using the RP, PR, and FR models, we established a correspondence between the theoretical framework and the observed QPO frequencies from seven X-ray binary sources: GRO J1655-40, XTE J1550-564, XTE J1859+226, GRS 1915+105, H1743-322, $M82 X_1$, and $Sgr A^{*}$. By applying the MCMC analysis, the likelihood evaluation is performed using a Python-based pipeline developed for the current BH model. The resulting posterior samples are analyzed with the \texttt{GetDist} package to extract marginalized constraints on each parameter and to generate one- and two-dimensional posterior distributions for the mass $m$, spin parameter $a/m$, and orbital radius $r/m$ at the 68\% confidence level (CL), while the accelerating parameter $m\cdot A$ is constrained at the 90\% confidence level. 

Our analysis shows that the inferred BH masses, spins, and orbital radii are consistent with the observational data, while the peak of the dimensionless accelerating parameter $m\cdot A$ is consistent with zero, and upper bounds are found within $0.003-0.020$ for all the sources. This implies that any plausible acceleration of the accelerating BH is too weak to affect the observed QPO timing at the existing sensitivity. Also, our result confirms that the frame-dragging effect is produced due to the spin parameter $a/m$, while the dimensionless accelerating parameter contributes only a weak secondary effect, and it is insufficient to alter the characteristic frequency ratios. Hence, we conclude that the spacetime around the accelerating BH remains effectively Kerr-like for all the analyzed sources. It is interesting to note that the accelerating BH parameters exhibit model dependence. In the RP model, the dimensionless accelerating parameter $m\cdot A$ shows a mild anti-correlation with the spin and mass parameters, while showing a positive correlation with the orbital radius $r/m$. Furthermore, the mass parameter is weakly correlated with the BH spin and strongly anti-correlated with the orbital radius $r/m$. These dependencies arise because higher spin or increased acceleration enhances the orbital frequencies and frame-dragging effect, which are partially compensated by decreasing the acceleration or increasing the orbital radius. In the PR model, the upper bounds of the parameter $m\cdot A$ are very small ($0.003$--$0.009$) for all seven X-ray binary sources, indicating that $m\cdot A$ shows negligible correlation with all other parameters. This implies that the acceleration has a negligible effect on the resonance dynamics. In the FR model, we find that the dimensionless parameter $m\cdot A$ exhibits stronger coupling, being anti-correlated with the mass and positively correlated with both the orbital radius and spin. The relatively higher upper limits ($0.015$--$0.020$) of $m\cdot A$ suggest that small variations in acceleration may influence the nonlinear coupling between the radial and vertical oscillations.

From this analysis, we conclude that the negligible values of the dimensionless parameter $m\cdot A$ inferred from the three QPO models—the RP, PR, and FR models—indicate that the accelerating Kerr spacetime of the seven observed X-ray binaries is consistent with the Kerr metric. The accelerating factor acts as a small perturbation in the strong gravitational field, leaving the essential dynamical features—frame dragging, disk precession, and resonance structure—unaltered. While BHs in dense astrophysical environments, such as globular clusters or in the vicinity of other massive objects, can experience a nonzero net acceleration, our results suggest that most of the analyzed X-ray binaries may reside in
isolated environments.

\renewcommand{\theequation}{5.\arabic{equation}} \setcounter{equation}{0}

\section*{Acknowledgements}

 This work is supported by the National Natural Science Foundation of China under Grants No.~12275238, the Zhejiang Provincial Natural Science Foundation of China under Grants No.~LR21A050001 and No.~LY20A050002, the National Key Research and Development Program of China under Grant No. 2020YFC2201503, and the Fundamental Research Funds for the Provincial Universities of Zhejiang in China under Grant No.~RF-A2019015. 
 
\appendix

\section{Appendix A: The expressions of three fundamental frequencies}
\renewcommand{\theequation}{A.\arabic{equation}} \setcounter{equation}{0}

For an accelerating BH described by the Kerr-C metric, the three fundamental frequencies $\nu_\phi$, $\nu_r$, and $\nu_\theta$ are given by
\begin{widetext}
\begin{eqnarray}
v_{\phi}=\frac{\Omega_{\phi}}{2\pi}&=&
\frac{r^2 \sqrt{m \left(A^2 r+\frac{1}{r}\right)-A^2 r^2}-a_{*} m \left(A^2 m r^2-A^2 r^3+m\right)}{r^3-a_{*}^2 m^2 \left(A^2 m r^2-A^2 r^3+m\right)},\label{a12}
\end{eqnarray}
\begin{eqnarray}
\nu_{r} &= -&\nu_{\phi}\Bigg[\Big(8 a_{*} m r(A^2 r^3-m (A^2 r^2+1)) \sqrt{m (A^2 r+\frac{1}{r})-A^2 r^2}+a_{*}^2 m^2(m(3 A^4 r^4+2 A^2 r^2+3)-4 A^4 r^5)\nonumber \\&&+r(4 A^2 r^4+m^2(-2 A^4 r^4+12 A^2 r^2+6)+m r(3 A^4 r^4-18 A^2 r^2-1))\Bigg]\Bigg[r^2 (A^2 m r^2-A^2 r^3+m)\Bigg]^{-1},\label{a19}
\end{eqnarray}
and 
\begin{eqnarray}
\nu_{\theta} &=& -\nu_{\phi}\Bigg[4 a_{*} m r(2 A^2 m^2 r-3 A^2 m r^2+A^2 r^3+m) \sqrt{m(A^2 r+\frac{1}{r})-A^2 r^2}+a_{*}^2 m^2 (-2 A^4 r^5-4 A^2 m^2 r+ \nonumber \\&&m(A^4 r^4+6 A^2 r^2-3))-r^2(-2 A^2 r^3+4 m^3(A^4 r^2+A^2)-4 m^2 (2 A^4 r^3+A^2 r)+m (3 A^4 r^4+4 A^2 r^2+1))\Bigg]\nonumber \\&&\Bigg[r^2(A^2 m r^2-A^2 r^3+m)\Bigg]^{-1},\label{a20}
\end{eqnarray} 
where $a_*\equiv a/J$.
\end{widetext}


\end{document}